\documentstyle[12pt]{article}

\input{tcilatex}

\begin{document}

\title{On the Dynamical Interpretation for the Quantum-Measurement
Projection Postulate}
\author{V. P. Belavkin \\
Mathematics Department, University of Nottingham,\\
NG7 2RD, UK\\
\and R. L. Stratonovich \\
Mathematics Department, University of Nottingham, \\
NG7 2RD, UK}
\date{Received June 11, 1996. Published in:\\
{\it Intern. Journal of Theoretical Physics} {\bf 35} (1996) No 11, 2215 -
2228.}
\maketitle

\begin{abstract}
An apparatus model with discrete momentum space suitable for the exact
solution of the problem is considered. The special Hamiltonian of its
interaction with the object system under consideration is chosen. In this
simple case it is easy to illustrate how difficulties in constructing the
dynamical interpretation of selective collapse could be overcome without any
limiting procedure. For this purpose one can apply either averaging with
respect to a non-quantum parameter or reducing the algebra of joint-system
operators (i. e. passing from algebra ${\cal A}$ of operators to a
subalgebra ${\cal A}_0$). The latter procedure implies averaging with
respect to apparatus quantum variables not belonging to ${\cal A}_0$.
\end{abstract}

\footnote{%
On leave of absence from Physics Department, Moscow State University, 119899
Moscow, Russia}

\section{Introduction}

In this paper we consider the dynamical interpretation of the selective
collapse in the one dimensional case, when momentum of the apparatus has
discrete spectrum of eigenvalues. This simplifies the problem of dynamical
corroboration of the von Neumann projection postulate. The idea to consider
the case, when one of two main conjugate dynamic variables (momentum or
coordinate) are discrete, and to take the apparatus state commuting with
discrete variable belongs to one of the authors [1].

The approach to the problem of the selective collapse interpretation is
quite ordinary and is well known from the time of von~Neumann [2]. The
collapse of the quantum object state, which takes place during measurement
of an object variable $X$ with discrete spectrum, is interpreted with the
help of interaction between object ${\rm S}$ and apparatus ${\rm A}$ (the
latter being in the quasiclassical state) and with the help of the
subsequent classical-like measurement of some apparatus variable $Y$. In our
case $Y$ depends on quantum momentum $\hat p$. Moreover, in this case the
evolution opertaor can exactly realize the transformation of the product
wave function 
\[
|\varphi\rangle\otimes |y_0\rangle = \sum_jc_j |x_j\rangle \otimes
|y_0\rangle 
\]
for the joint system into the correlated one 
\[
\sum_jc_j|x_j\rangle\otimes |y_j\rangle. 
\]
This transformation was proposed by von Neumann for the measurement
interpretation. Here $|x_j\rangle$ are the eigenfunctions of $X$, and $|y_j
\rangle$ are the eigenfunctions of $Y$. In contrast to the von Neumann
theory, use of the mixed apparatus state or to be exact the quasi-classical
state is desirable because eigenvalues $\{ y_j\}$ of $Y$ can only be
distinguished from one another macroscopically in a quasi-classical state,
the appropriate measured operator $Y$ being chosen. Moreover we use an
averaging procedure of the apparatus state (see [3]). This procedure helps
to overcome the difficulties connected with the dynamic interpretation of
the collapse making the apparatus state compatible with $Y$.

Our goal is to interprete the selective collapse 
$$
\rho_{{\rm S}}\rightarrow\frac{1}{w_l} E_l\rho_{{\rm S}}E_l \eqno (1.1) 
$$
($w_l = {\rm Tr}_{{\rm S}}\rho_{{\rm S}}E_l)$ of the density matrix of the
quantum object S. According to the projection postulate it takes place when
the result $x_l$ of measurement of the operator $X = \sum_j x_jE_j$ becomes
known. Here $E_j$ are the orthogonal projectors ($\: E_jE_k = E_j\delta_{jk}$%
, $\:\sum_jE_j = I_{{\rm S} }$).

Our treatment is restricted to the following assumptions.

(i) The coordinate space of the apparatus model is finite, namely it is of
the length $L$ and is curved into itself (like circumference), the
coordinate spectrum being, say, the interval $[-L/2, L/2]$. This means that
the shift $q\rightarrow q + a$ gives $q + a - L$ if $L/2 < q + a< 3L/2$ and $%
q + a + L$ if $-L/2 > q + a>- 3L/2$. The pointer on a fixed axis (for it $q=
\varphi$ is the angle, $L = 2\pi$) or a box with periodic boundary
conditions may serve as examples. For an arbitrary $L$ the apparatus
momentum has eigenvalues $p_k = 2\pi\hbar k/L$.

(ii) The initial apparatus density matrix $\rho _{{\rm A}}$ of the apparatus
is compatible with momentum $\hat{p}$, i. e. is diagonal in the momentum
representation 
$$
\rho _{kl}:=\langle p_{k}|\rho _{{\rm A}}|p_{l}\rangle =w_{k}^{0}\delta
_{kl}.\eqno(1.2)
$$%
Besides we suppose that 
$$
w_{k}^{0}=0\quad {\rm at}\quad |k|>m.\eqno(1.3)
$$%
This compatible density matrix is only possible because of discrete
character of the momentum spectrum. In fact, its continuous variant 
\[
\langle p^{\prime }|\rho _{{\rm A}}|p\rangle =w^{0}(p)\delta (p^{\prime }-p)
\]%
is impossible because this operator has infinite trace (if $w^{0}(p)$ is not
equal to zero everywhere).

(iii) The interaction Hamiltonian is of the form 
$$
{\cal H}_{{\rm int}}(t)=-B\otimes (\gamma \hat{q}+\lambda I_{{\rm A}})\delta
(t),\eqno(1.4)
$$%
where $\gamma $, $\lambda $ are interaction constants, $I_{{\rm A}}$ is the
apparatus identity operator. Of course, the presence of delta-function on
the right-hand side of (1.4) makes the process of interaction somewhat
unrealistic. This delta-function type of interaction was applied in [4] in
the recurrent variant for realizing continuous observation.

The operator 
$$
B=f(X)=\sum_jb_jE_j=\sum_jf(x_j)E_j \eqno (1.5) 
$$
enters the right-hand side of (1.4), the function $f$ being chosen in such a
way that all eigenvalues $b_j$ of $B$ be multiple to the same quantity $a >
0 $: 
$$
b_j = n_ja. \eqno (1.6) 
$$
Here $n_j$ are integers that increase with increasing $j$. Transformation $%
b_j=f(x_j)$ is supposed to be non-degenerate. The necessety of (1.6) will be
clear later.

To obtain the collapse (1.1) of the object system state, the measurement of
the variable $Y$ depending on the apparatus momentum will be made. The
matrix density (1.2) is very convenient for measuring $Y$ because it
commutes with $p$ and therefore with $Y(\hat p)$.

In the general case the selective quantum collapse 
$$
\rho _{{\rm A}}\rightarrow \frac{1}{w_{l}^{\prime }}P_{l}\rho _{{\rm A}}P_{l}%
\eqno(1.7)
$$%
of the apparatus state takes place after measurement of $Y=\sum_{j}y_{j}P_{j}
$ if the measurement result $y_{l}$ becomes known. Here $P_{j}$ are
eigen-orthoprojectors of $Y$ ($\sum_{j}P_{j}=I_{{\rm A}}$) and $%
w_{l}^{\prime }={\rm Tr}_{{\rm A}}P_{l}\rho _{{\rm A}}$. Averaging $%
\sum_{l}w_{l}^{\prime }\tilde{\rho}_{l}$ {\it a\ posteriori} matrix
densities 
\[
\tilde{\rho}_{l}=\frac{1}{w_{l}^{\prime }}P_{l}\rho _{{\rm A}}P_{l}
\]%
does not give {\it a priori} matrix $\rho _{{\rm A}}$ in the general case.
This means that the condition of consistency 
$$
\sum_{l}w_{l}^{\prime }\tilde{\rho}_{l}=\rho _{{\rm A}}\quad {\rm or}\quad
\sum_{l}P_{l}\rho _{{\rm A}}P_{l}=\rho _{{\rm A}}\eqno(1.8)
$$%
is not obliged to be met. In our case the projectors $P_{j}$ defined by 
$$
P_{j}=\vartheta _{j}(\hat{p})\eqno(1.9)
$$%
commute with $\rho _{{\rm A}}$ and therefore the consistency condition (1.8)
is met. The function $\vartheta _{j}(\xi )$ is defined by (4.6).

As was pointed out in [3], the quasi-classical collapse 
$$
\rho _{{\rm A}}\rightarrow \frac{1}{w_{l}^{\prime }}\rho _{{\rm A}}\ast P_{l}%
\eqno(1.10)
$$%
obviously meeting the consistency condition can be applied in some cases.
Here the operation $\ast $ is defined with the help of the Wigner
transformation (2.7), (2.8) denoted by ${\cal W}$. To be exact, in our case 
\[
A\ast B=L\,{\cal W}^{-1}\{{\cal W}[A]{\cal W}[B]\}.
\]%
For projectors (1.9) we have 
\[
L\,{\cal W}[\vartheta _{l}(\hat{p})]=\vartheta _{l}(p_{j}),
\]%
and (1.10) is equivalent to 
\[
\rho _{{\rm A}}\rightarrow (w_{l}^{\prime })^{-1}{\cal W}^{-1}[{\cal W}[\rho
_{{\rm A}}]\vartheta _{l}(p_{j})]
\]%
or if we apply ${\cal W}$ to both sides of the formula 
$$
w(q,p_{j})\rightarrow \frac{1}{w_{l}^{\prime }}w(q,p_{j})\vartheta
_{l}(p_{j})\eqno(1.11)
$$%
This is nothing else as transition to the conditional distribution, which is
well-known non-quantum procedure. Using (2.10), one can easily see that
collapse (1.10), (1.11) is exactly equivalent to (1.7) in our simple case.
Because of this fact and because the condition (1.8) is met in our case, we
call the measurement of $Y=\sum y_{k}\vartheta _{k}(p)$ classical-like.

\section{The initial apparatus state in other representations}

Eigenfunctions of momentum $\hat{p}$ corresponding to the eigenvalues $%
p_{k}=2\pi \hbar k/L$ are 
$$
\psi _{k}(q)=L^{-1/2}\exp ({\rm i}p_{k}q/\hbar ).\eqno(2.1)
$$%
(the coordinate representation). Using expression on the right-hand side
taken at various $k$ we readily write down the matrix elements 
$$
V_{qk}=L^{-1/2}\exp (2\pi {\rm i}kq/L)\eqno(2.2)
$$%
of the unitary operator $V$ that transforms $\hat{p}$-representation to $%
\hat{q}$-representation and vice versa. Thus $\hat{q}$-representation of the
density matrix is 
\[
\rho (q^{\prime },q):=\langle q^{\prime }|\rho _{{\rm A}}|q\rangle
=\sum_{kl}V_{q^{\prime }k}\rho _{kl}V_{lq}^{\dagger },
\]%
or due to (1.2) and (2.2) 
$$
\rho (q^{\prime },q)=L^{-1}\sum_{k}\exp [2\pi {\rm i}(q^{\prime
}-q)k/L]w_{k}^{0}\eqno(2.3)
$$%
Therefore the coordinate probability density $w^{0}(q)=\rho (q,q)$ is
uniform 
\[
w^{0}(q)=1/L.
\]%
Hence we find the coordinate mean $\langle q\rangle =0$ and mean square 
$$
\sigma _{q}^{2}:=\langle q^{2}\rangle =\frac{1}{L}\int_{-L/2}^{L/2}q^{2}{\rm %
d}q=\frac{L^{2}}{12}.\eqno(2.4)
$$%
On the other hand the momentum mean square is 
$$
\sigma _{p}^{2}=\sum_{k=-m}^{m}p_{k}^{2}w_{k}^{0}=8\pi ^{2}\hbar
^{2}L^{-2}\sum_{k=1}^{m}k^{2}w_{k}^{0}\eqno(2.5)
$$%
if $w_{-k}^{0}=w_{k}^{0}$. According to (2.4), (2.5) we have 
$$
\sigma _{q}^{2}\sigma _{p}^{2}=\frac{2}{3}\pi ^{2}\hbar
^{2}\sum_{k=1}^{m}k^{2}w_{k}^{0}.\eqno(2.6)
$$%
It should be noted that we get $\,\sigma _{q}\sigma _{p}=0\,$ from (2.6) if $%
\,m=0$, i.e. if $\,w_{k}^{0}=\delta _{k0}$. This equation is very unusual
since it violates the Heisenberg uncertainty relation $\sigma _{q}\sigma
_{p}\geq \hbar /2$. Possibility of this paradox is argumented in Appendix.

When $\sigma_q\sigma_p\gg\hbar$, the apparatus is in a quasi-classical
state. We will suppose that this inequality is valid because the direct
macroscopic observation of a physical quantity is possible only in this
case. Owing to (1.3) and normalization condition $\sum_kw_k^0 = 1$, the
inequality $m\gg 1$ is a necessary condition for $\sigma_q\sigma_p\gg\hbar$.
For many distributions, i.e. for the uniform one formula $m\gg 1$ is also a
sufficient condition of a quasi-classical state.

Another representation of the apparatus state is the Wigner distribution,
which in our case takes the form 
\[
w(q,p_{j})=\frac{1}{L}\int_{-L/2}^{L/2}\exp \Big(-\frac{{\rm i}}{\hbar }%
up_{j}\Big)\rho \Big(q+\frac{u}{2},q-\frac{u}{2}\Big){\rm d}u
\]%
$$
=\frac{1}{L}\sum_{k=-\infty }^{\infty }\,\sum_{l=-\infty }^{\infty }\exp %
\Big[\frac{{\rm i}}{\hbar }q(p_{k}-p_{l})\Big]\Delta \Big(\frac{k+l}{2}-j%
\Big)\rho _{kl}.\eqno(2.7)
$$%
Here $\Delta (\eta )=\int_{-1/2}^{1/2}\exp {(2\pi {\rm i}\eta v)}dv$, i. e. 
\[
\Delta (\eta )=\frac{\sin (\pi \eta )}{\pi \eta }=%
\cases{\delta_n & at
$\eta = n$, \cr (-1)^n\pi^{-1}/(n + \mbox{\small ${1\over 2}$}) & at
$\eta = n + \mbox{\small ${1\over 2}$}$\cr}
\]%
($n$ is integer). We denote transformation (2.7) by ${\cal W}$; 
$$
{\cal W}[\rho _{{\rm A}}]=w(q,p_{j}).\eqno(2.8)
$$%
It is easy to check that $w(q,p_{j})$ has properties 
\[
\sum_{j}w(q,p_{j})=\rho (q,q),\quad \int w(q,p_{j}){\rm d}q=\rho
_{jj}=w_{j}^{0}
\]%
usual for the Wigner distribution. Moreover the formula 
$$
{\rm Tr_{A}}\,G\rho _{{\rm A}}=L\sum_{j=-\infty }^{\infty }\int_{-L/2}^{L/2}%
{\cal W}[G]{\cal W}[\rho _{{\rm A}}]{\rm d}q\eqno(2.9)
$$%
is valid. For the special matrix density (1.2) we get 
$$
w(q,p_{j})=w_{j}^{0}/L.\eqno(2.10)
$$

\section{Interaction between the object system ${\rm S}$ and apparatus}

Let $H_{{\rm S}}$ be a Hamiltonian acting in the Hilbert space ${\cal H}_{%
{\rm S}}$ of the object system ${\rm S}$. The apparatus Hamiltonian $H_{{\rm %
A}}$ is an operator acting in ${\cal H}_{{\rm A}}$. Interaction between $%
{\rm S}$ and ${\rm A}$ that lasts very short time from $t=-\varepsilon $ to $%
t=\varepsilon $ is described by the interaction Hamiltonian (1.4) acting on $%
{\cal H}_{{\rm S}}\otimes {\cal H}_{{\rm A}}$, $B$ being the ${\rm S}$%
-system operator with discrete eigenvalues (1.6). Its measurement or ---
what is equivalent --- measurement of $X$ is to be interpreted. Hence the
total Hamiltonian assumes the form 
$$
H(t)=H_{{\rm S}}\otimes I_{{\rm A}}+I_{{\rm S}}\otimes H_{{\rm A}}-\gamma
B\otimes q\delta (t)-\lambda B\otimes I_{{\rm A}}\delta (t).\eqno(3.1)
$$%
The state of the joint system ${\rm S}+{\rm A}$ at the initial instant $%
t_{0}=-\varepsilon $ is given by the density matrix 
$$
\rho (-\varepsilon )=\rho _{{\rm S}}\otimes \rho _{{\rm A}}.\eqno(3.2)
$$%
In the Schr\"{o}dinger picture the density matrix depends on time as 
$$
\rho (t)=U(t,t_{0})\rho (t_{0})U^{\dagger }(t,t_{0}),\eqno(3.3)
$$%
where the evolution operator $U$ is given by 
$$
U(t,t_{0})={\cal T}\exp \Big[-\frac{{\rm i}}{\hbar }\int_{t_{0}}^{t_{1}}%
\!H(t){\rm d}t\Big].\eqno(3.4)
$$%
Here ${\cal T}$ denotes the time ordering of operators $H(t)$, namely the
greater $t$ is the more to the left $H(t)$ stands. We choose $%
t_{1}=\varepsilon >0$, where $\varepsilon $ is a very small number. Then
(3.3) gives 
$$
\rho (\varepsilon )=\exp \Big[\frac{{\rm i}}{\hbar }B\otimes (\gamma \hat{q}%
+\lambda I_{{\rm A}})\Big](\rho _{{\rm S}}\otimes \rho _{{\rm A}})\exp \Big[-%
\frac{{\rm i}}{\hbar }B\otimes (\gamma \hat{q}+\lambda I_{{\rm A}})\Big]\eqno%
(3.5)
$$%
owing (3.1), (3.2), (3.4). We will use the orthogonal projectors $\{E_{j}\}$
corresponding to the operator $B=\sum_{j}b_{j}E_{j}$. As is well known, for
them 
$$
I_{{\rm S}}=\sum_{j}E_{j}.\eqno(3.6)
$$%
By virtue of (3.6) we can take $\sum_{i}E_{i}\rho _{{\rm S}}\sum_{j}E_{j}$
instead of $\rho _{{\rm S}}$ in (3.5) and obtain 
$$
\rho (\varepsilon )=\sum_{ij}\exp \Big[\frac{{\rm i}}{\hbar }B\otimes
(\gamma \hat{q}+\lambda I_{{\rm A}})\Big](E_{i}\rho _{{\rm S}}E_{j}\otimes
\rho _{{\rm A}})\exp \Big[-\frac{{\rm i}}{\hbar }B\otimes (\gamma \hat{q}%
+\lambda I_{{\rm A}})\Big].\eqno(3.7)
$$%
But $BE_{i}=b_{i}E_{i}$, $E_{j}B=E_{j}b_{j}$ and $g(B\otimes D)E_{j}\otimes
\rho _{{\rm A}}=E_{j}\otimes (g(b_{j}D)\rho _{{\rm A}})$ for an arbitrary $c$%
-function $g$. Therefore (3.7) yields 
$$
\rho (\varepsilon )=\sum_{ij}E_{i}\rho _{{\rm S}}E_{j}\otimes \exp \Big[%
\frac{{\rm i}}{\hbar }b_{i}(\gamma \hat{q}+I_{{\rm A}})\Big]\rho _{{\rm A}%
}\exp \Big[-\frac{{\rm i}}{\hbar }b_{j}(\gamma \hat{q}+I_{{\rm A}})\Big].%
\eqno(3.8)
$$%
Now we use formulas (1.4) and let 
\[
a\gamma =2\pi \hbar (2m+1)/L.
\]%
Then in the apparatus coordinate representation 
\[
\langle q^{\prime }|\rho (\varepsilon )|q\rangle =\sum_{ij}E_{i}\rho _{{\rm S%
}}E_{j}e^{{\rm i}(n_{i}-n_{j})\chi }\exp \Big[2\pi {\rm i}%
(2m+1)(n_{i}q^{\prime }-n_{j}q)L^{-1}\Big]\rho (q^{\prime },q)
\]%
with $\chi =a\lambda /\hbar $. Substituting (2.3) into the right-hand side
and passing to the $p$-representation hence we get 
$$
\langle p_{r}|\rho (\varepsilon )|p_{s}\rangle =\sum_{ij}E_{i}\rho _{{\rm S}%
}E_{j}e^{{\rm i}(n_{i}-n_{j})\chi }w_{r-(2m+1)n_{i}}^{0}\delta
_{r-s-(2m+1)(n_{i}-n_{j})}.\eqno(3.9)
$$%
The following Wigner transform follows from this result 
\[
{\cal W}[\rho (\varepsilon )]_{q,p_{k}}=\sum_{ij}E_{i}\rho _{{\rm S}}E_{j}e^{%
{\rm i}(n_{i}-n_{j})\chi }\exp \Big[2\pi {\rm i}(2m+1)(n_{i}-n_{j})\frac{q}{L%
}\Big]
\]%
$$
\qquad \qquad \qquad \times w\Big(q,p_{k}-\mbox{\small ${1\over 2}$}%
(p_{(2m+1)n_{i}}+p_{(2m+1)n_{j}})\Big),\eqno(3.10)
$$%
if all $n_{i}+n_{j}$ are even.

\section{The apparatus physical quantity that should be measured}

Let us consider the expression 
$$
R(p_{r}):=\langle p_{r}|\rho (\varepsilon )|p_{r}\rangle ,\eqno(4.1)
$$%
which in our case, due to (3.9), assumes the form 
$$
R(p_{r})=\sum_{j}E_{j}\rho _{{\rm S}}E_{j}w_{r-(2m+1)n_{j}}^{0}.\eqno(4.2)
$$%
It is an operator on ${\cal H}_{{\rm S}}$ and simultaneously the
distribution of momentum $p_{j}$. We see that correlation exists between
values $b_{j}$ of $B$ and those of $p$. In fact, the density matrix 
$$
\tilde{\rho}_{j}=\frac{1}{w_{j}}E_{j}\rho _{{\rm S}}E_{j},\eqno(4.3)
$$%
in which $B$ has definite value $b_{j}$, enters the same term $w_{j}\tilde{%
\rho}_{j}w_{r-(2m+1)n_{j}}^{0}$ of the sum (4.2) as distribution $%
w_{r-(2m+1)n_{j}}^{0}$ which is not 0 in the range 
\[
-2\pi \hbar m/L\leq p_{r}-2\pi \hbar (2m+1)n_{j}/L\leq 2\pi \hbar m/L
\]%
(according to (1.3)) i.e. 
$$
2\pi \hbar \lbrack (2m+1)n_{j}-m]/L\leq p_{r}\leq 2\pi \hbar \lbrack
(2m+1)n_{j}+m]/L.\eqno(4.4)
$$%
Therefore determining the range, to which the momentum belongs, signifies
determining the value of $B$ and $X$. Let us denote the range (4.4) by $S_{j}
$. Thus 
$$
w_{r-(2m+1)n_{j}}^{0}=%
\cases{w_{r-(2m+1)n_j}^0 & at $p_r\in S_j$,\cr
0 & at $p_r\notin S_j$.\cr }\eqno(4.5)
$$%
Various ranges never overlap because $n_{j+1}-n_{j}\geq 1$. Let us take the
enlarged not overlapping ranges $\tilde{S}_{j}$ such that each $\tilde{S}_{j}
$ includes $S_{j}$ and so that the sum $\sum_{j}\tilde{S}_{j}$ is equal to
the set of all $p_{j}$, $j=0,\pm 1,\pm 2,\ldots $. This enlarging can be
made in various ways. For example, we can take the points 
$$
s_{j}=2\pi \hbar L^{-1}\Big[(2m+1)\frac{n_{j}+n_{j+1}}{2}\Big]_{{\rm IP}}%
\eqno(4.6)
$$%
(the subscript IP means the integral part) lying approximately on the
half-way between $S_{j}$ and $S_{j+1}$ and define $\tilde{S}_{j}$ as the
range $s_{j-1}<p_{k}\leq s_{j}$. Now we define the function 
$$
\vartheta _{j}(p_{k})=%
\cases{1 & at $p_k\in\tilde S_j$,\cr 0 
& otherwise.\cr }\eqno(4.7)
$$%
From (4.5), (4.7) and since $S_{j}$ is the subset of $\tilde{S}_{j}$, we
have 
$$
w_{k-(2m+1)n_{i}}^{0}\vartheta _{j}(p_{k})=w_{k-(2m+1)n_{i}}^{0}\delta _{ij}.%
\eqno(4.8)
$$%
Let the measured apparatus operator be 
$$
Y(\hat{p})=\sum_{j}p_{(2m+1)n_{j}}\vartheta _{j}(\hat{p})\eqno(4.9)
$$%
($p_{(2m+1)n_{j}}$ being the central point of $S_{j}$), or 
$$
Y=\sum_{j}j\vartheta _{j}(\hat{p}).\eqno(4.10)
$$%
The equation (4.9) corresponds to inexact measurement of $\hat{p}$, the
latter one means that the number $j$ of range, to which $p$ belongs, is
measured. Note that we may set $Y=\sum_{j}x_{j}\vartheta _{j}(\hat{p})$,
then we will have $\langle \lbrack X\otimes I_{{\rm A}}-I_{{\rm S}}\otimes
Y]^{2}\rangle =0$ as it follows from (5.5), (4.2), (4.8).

\section{Selective collapse of the ${\rm S}$-system state as a result of
measuring apparatus variable $Y$}

Now if we measure the physical quantity (4.7) or (4.8) and $p$ proves to
belong to $\tilde{S}_{l}$, the collapse 
$$
\rho (\varepsilon )\rightarrow \frac{1}{w_{l}^{\prime }}\Big[I_{{\rm S}%
}\otimes \vartheta _{l}(\hat{p})\Big]\rho (\varepsilon )\Big[I_{{\rm S}%
}\otimes \vartheta _{l}(\hat{p})\Big]\eqno(5.1)
$$%
(with $w_{l}^{\prime }={\rm Tr}\,[I_{{\rm S}}\otimes \vartheta _{l}(\hat{p}%
)]\rho (\varepsilon )[I_{{\rm S}}\otimes \vartheta _{l}(\hat{p})]$) takes
the form 
$$
\langle p_{r}|\rho (\varepsilon )|p_{s}\rangle \rightarrow \frac{1}{%
w_{l}^{\prime }}\vartheta _{l}(p_{r})\langle p_{r}|\rho (r)|p_{s}\rangle
\vartheta _{l}(p_{s})=\frac{1}{w_{l}^{\prime }}E_{l}\rho _{{\rm S}%
}E_{l}w_{s-(2m+1)n_{l}}^{0}\delta _{rs}.\eqno(5.2)
$$%
owing to (3.9), (4.8). In fact, applying (4.8) we have 
\[
\vartheta _{l}(p_{r})w_{r-(2m+1)n_{i}}^{0}\delta
_{r-s-(2m+1)(n_{i}-n_{j})}=w_{r-(2m+1)n_{i}}^{0}\delta
_{r-s-(2m+1)(n_{i}-n_{j})}\delta _{il}
\]%
\[
\qquad \qquad \qquad \qquad =w_{s-(2m+1)n_{j}}^{0}\delta
_{r-s-(2m+1)(n_{i}-n_{j})}\delta _{il}
\]%
and 
\[
w_{s-(2m+1)n_{j}}^{0}\vartheta _{l}(p_{s})=w_{s-(2m+1)n_{l}}^{0}\delta _{jl}.
\]%
This leads to (5.2). Formula (5.2) means that the {\it a posteriori} state
of quantum object ${\rm S}$ is $E_{l}\rho _{{\rm S}}E_{l}/w_{l}^{\prime }$ = 
$E_{l}\rho _{{\rm S}}E_{l}/w_{l}$.

However, the objection arises that it is incorrect to interprete the quantum
collapse $\rho _{{\rm S}}\rightarrow E_{l}\rho _{{\rm S}}E_{l}/w_{l}$ by
another quantum collapse, namely by (5.1). In fact, matrix (3.9) does not
commute with $I_{{\rm S}}\otimes \hat{p}$ and $\{I_{{\rm S}}\otimes Y\}$ and
therefore consistency condition of the type (1.8) is violated. This
condition would had been met for collapse 
$$
\rho (\varepsilon )\rightarrow \frac{1}{w_{l}^{\prime }}\rho (\varepsilon
)\ast \vartheta _{l}(\hat{p}),\eqno(5.3)
$$%
but now (5.3) is not justified since it contradicts the collapse (5.1).

To overcome the above difficulty, the averaging with respect to some quantum
or non-quantum variables should be done. There exist several lines of action
and reasoning.

{\bf 1}. We suppose that non-quantum parameter $\chi $ entering the
right-hand side of (3.10) is random and uniformly distributed on the
interval $-\pi <\chi \leq \pi $. Then everaging the right-hand side of
(3.10) with respect to $\chi $ leads to 
$$
\overline{\langle p_{r}|\rho (\varepsilon )|p_{s}\rangle }=\sum_{j}E_{j}\rho
_{{\rm S}}E_{j}\,w_{r-(2m+1)n_{j}}^{0}\delta _{rs}\eqno(5.4)
$$%
because the mean value of $\exp [{\rm i}(n_{i}-n_{j})\chi ]$ is $\delta _{ij}
$. The matrix density (5.4) commutes with $I_{{\rm S}}\otimes \hat{p}$ and $%
I_{{\rm S}}\otimes Y(\hat{p})$. Therefore the measurement of $Y$ is
classical-like (see Sect.1) and both the quantum collapse (5.1) and the
classical one (5.3) may now be applied to (5.4). This gives the resulting 
{\it a~posteriori} state $E_{l}\rho _{{\rm S}}E_{l}w_{r-(2m+1)n_{l}}^{0}%
\delta _{rs}/w_{l}^{\prime }$. Averaging with respect to the apparatus
parameter was used in [6] for explaining the non-selective collapse.

{\bf 2}. Another possibility is the averaging with respect to some quantum
variables of the apparatus. We can restrict the operator algebra in which we
are interested in. Let us only consider operator subalgebra ${\cal A}_{0}$
generated by all operators of ${\rm S}$-system (i.e. operators of the type $%
D\otimes I_{{\rm A}}$) and by operator $I_{{\rm S}}\otimes \hat{p}$. The
analogous type of the operator subalgebra (with coordinate taken instead of
momentum) was considered in [1]. To be exact algebra of all operators
commuting with $Q=\kappa I\otimes \hat{q}$ was applied there for securing
the consistancy condition by defining non-demolition observation continuous
in time, the operator $Y$ having both discreate and continuous specrtrum.
Earlier Araki [5] used a special subalgebra of operators for obtaining
non-selective collapse in the limit $t\rightarrow \infty $ for a particular
choice of interaction.

The state functional (functional of mean values) for operators belonging to
our subalgebra ${\cal A}_0$ is defined with the help of operator (4.1):

$$
\langle G\rangle =\sum_{k}{\rm Tr}_{{\rm S}}\,R(p_{k})\langle
p_{k}|G|p_{k}\rangle ,\eqno(5.5)
$$%
When we only consider operators from the subalgebra ${\cal A}_{0}$ and use $%
R(p_{k})$, the classical selective collapse 
$$
R(p_{k})\rightarrow \frac{1}{w_{l}^{\prime }}R(p_{k})\vartheta _{l}(p_{k})%
\eqno(5.6)
$$%
analogous to transition to the conditional probability distribution takes
place provided that the result of the measurement becomes known. According
to (4.2), (4.8) this means transformation 
\[
R(p_{k})\rightarrow \frac{1}{w_{l}^{\prime }}E_{l}\rho _{{\rm S}%
}E_{l}w_{k-(2m+1)n_{l}}^{0}
\]%
Summation with respect to apparatus momentum gives {\it a~posteriori} state $%
E_{l}\rho _{{\rm S}}E_{l}/w_{l}^{\prime }$ of the quantum object.

{\bf 3}. Suppose now that the quantum system interacts with two systems $%
{\rm A}$ and ${\rm C}$, ${\rm C}$ being another copy of ${\rm A}$-system
considered earlier. Let it be in the same initial state. Then ${\rm A}+{\rm C%
}$ constitute a new complex apparatus. Averaging with respect to the
C-system variables, i. e. considering subalgebra ${\cal A}_{0}$ operators of
the type $D\otimes I_{{\rm C}}$ ($D$ being an operator on ${\cal H}_{{\rm S}%
}\otimes {\cal H}_{{\rm A}}$) will solve the problem. For operators $\tilde{D%
}=D\otimes I{\rm C}$ from ${\cal A}_{0}$ the functional of mean values is $%
\langle \tilde{D}\rangle ={\rm Tr}_{{\rm S}+{\rm A}}D\rho _{{\rm S}+{\rm A}}$
with $\rho _{{\rm S}+{\rm A}}={\rm Tr}_{{\rm C}}\rho $.

Now the total Hamiltonian takes the form 
\[
H(t)=H_{{\rm S}}^{\prime \prime }+H_{{\rm A}}^{\prime \prime }+H_{{\rm C}%
}^{\prime \prime }-\gamma B^{\prime \prime }(\hat{q}^{\prime \prime
}+Q^{\prime \prime })\delta (t),
\]%
where $H_{{\rm S}}^{\prime \prime }=H_{{\rm S}}\otimes I_{{\rm A}}\otimes I_{%
{\rm C}}$, $B^{\prime \prime }=B\otimes I_{{\rm A}}\otimes I_{{\rm C}}$, $%
\hat{q}^{\prime \prime }=I_{{\rm S}}\otimes \hat{q}\otimes I_{{\rm C}}=I_{%
{\rm S}}\otimes \hat{q}^{\prime }$, $Q^{\prime \prime }=I_{{\rm S}}\otimes
I_{{\rm A}}\otimes Q=I_{{\rm S}}\otimes Q^{\prime }$ and so on, $Q$ being
the coordinate of ${\rm C}$, i.e. the operator on ${\cal H}_{{\rm C}}$.
Naturally the matrix 
\[
\rho =\rho _{{\rm S}}\otimes \rho _{{\rm A}}\otimes \rho _{{\rm C}}
\]%
serves as the initial density matrix. In this case we have 
$$
\rho (\varepsilon )=\sum_{ij}E_{i}\rho _{{\rm S}}E_{j}\otimes \exp \Big[%
\frac{{\rm i}}{\hbar }\gamma b_{i}(\hat{q}^{\prime }+Q^{\prime })\Big](\rho
_{{\rm A}}\otimes \rho _{{\rm C}})\exp \Big[-\frac{{\rm i}}{\hbar }\gamma
b_{j}(\hat{q}^{\prime }+Q^{\prime })\Big]\eqno(5.7)
$$%
instead of (3.9). Since $\hat{q}^{\prime }$ commutes with $Q^{\prime }$ and $%
I_{{\rm A}}\otimes \rho _{{\rm C}}$, and $Q^{\prime }$ commutes with $\rho _{%
{\rm A}}\otimes I_{{\rm C}}$, this formula can be written as 
$$
\rho (\varepsilon )=\sum_{ij}E_{i}\rho _{{\rm S}}E_{j}\otimes e^{{\rm i}%
\gamma b_{i}\hat{q}/\hbar }\rho _{{\rm A}}e^{-{\rm i}\gamma b_{j}\hat{q}%
/\hbar }\otimes e^{{\rm i}\gamma b_{i}Q/\hbar }\rho _{{\rm C}}e^{-{\rm i}%
\gamma b_{j}Q/\hbar }.\eqno(5.8)
$$%
If we write the matrices $r_{ij}=\exp ({\rm i}\gamma b_{i}Q/\hbar )\rho _{%
{\rm C}}\exp (-{\rm i}\gamma b_{j}Q/\hbar )$ in the coordinate
representation, after using (1.3) we have 
$$
r_{ij}(Q^{\prime },Q)=\exp {[{\rm i}\hbar ^{-1}\gamma a(n_{i}Q^{\prime
}-n_{j}Q)]}\rho _{{\rm C}}(Q^{\prime },Q),\eqno(5.9)
$$%
where 
$$
\rho _{{\rm C}}(Q^{\prime },Q)=L^{-1}\sum_{k=-m}^{m}\exp [2\pi {\rm i}%
(Q^{\prime }-Q)k/L]w_{k}^{0}\eqno(5.10)
$$%
((5.10) is analogous to (2.3)). From (5.9), (5.10) we see that setting $%
\gamma a=2\pi \hbar N/L$ ($N$ is an integer) and taking the partial trace $%
{\rm Tr}_{{\rm C}}$ with respect to ${\rm C}$-system (i.e. integrating with
respect to $Q^{\prime }=Q$) will give 
\[
{\rm Tr}_{{\rm C}}\,r_{ij}=\delta _{ij}.
\]%
Therefore we get from (5.8) 
\[
{\rm Tr}_{{\rm C}}\,\rho (\varepsilon )=\sum_{j}E_{j}\rho _{{\rm S}%
}E_{j}\otimes \exp ({\rm i}\gamma an_{j}\hat{q}/\hbar )\rho _{{\rm A}}\exp (-%
{\rm i}\gamma an_{j}\hat{q}/\hbar )
\]%
and 
\[
\langle p_{k}|{\rm Tr}_{{\rm C}}\,\rho (\varepsilon )|p_{l}\rangle
=\sum_{j}E_{j}\rho _{{\rm S}}E_{j}w_{k-(2m+1)n_{j}}^{0}\delta _{kl}
\]%
for $N=2m+1$. Thus averaging with respect to all ${\rm C}$-system quantum
variables gives the same result as averaging in non-quantum random parameter 
$\chi $.

In the first and the third ways of reasoning we have obtained the {\it a
posteriori} combined system state $\tilde{\rho}_{l}^{({\rm S})}\otimes 
\tilde{\rho}_{l}^{({\rm A})}$, where $\tilde{\rho}_{l}^{({\rm S})}=E_{l}\rho
_{{\rm S}}E_{l}/w_{l}^{\prime }$, $\tilde{\rho}_{l}^{({\rm A}%
)}=\sum_{k}|p_{k}\rangle w_{k-(2m+1)n_{l}}^{0}\langle p_{k}|$. This means
that the quantum object is in the state $E_{l}\rho _{{\rm S}%
}E_{l}/w_{l}^{\prime }$. Due to normalization condition $w_{l}^{\prime }$
coinsides with the probability $w_{l}={\rm Tr}_{{\rm S}}\,E_{l}\rho _{{\rm S}%
}$ entering (1.1). So the transformation (1.1) of the object state takes
place.

\section*{Acknowledgments}

One of the authors (R. L. Stratonovich) is grateful to the Department of
Mathematics, University of Nottingham for its hospitality. The financial
support from E.P.S.R.C. (project GR/K08024) is gratefully acknowleged.

\section*{Appendix. Violation of the Heisenberg uncertainty relation?}

The Heisenberg uncertainty relation $\sigma_q\sigma_p \geq\hbar /2$ is the
consequence of the well-known operator inequality 
$$
4\langle A^2\rangle\langle B^2\rangle\geq \Big({\rm i}\langle [A,B]\rangle %
\Big)^2 \eqno ({\rm A}.1) 
$$
valid for any self-adjoint operators $A$ and $B$. Of course, it should be
valid in our case.

Let us map our coordinate space onto real axis in such a way that all points 
$x_n = x_0 + nL$ are images of the same coordinate-space point. Here $n$ is
an arbitrary interger. All functions on the coordinate space should appear
as periodical functions on the real axis. The momentum operator $p = -{\rm i}%
\hbar \partial /\partial x$ generates shifts 
\[
\exp ({\rm i} cp)\varphi (x) = \exp (c\hbar\partial /\partial x) \varphi (x)
= \varphi (x + \hbar c) 
\]
in the real axis and coordinate space. The normalized eigenfunctions of $p$
have the form 
$$
\varphi_k(x) = L^{-1/2}\exp ({\rm i} p_k x/\hbar ). \eqno ({\rm A}.2) 
$$
They correspond to eigenvalues $p_k = 2\pi\hbar k/L$.

Now the question arises how to define the function $q(M)$ in the coordinate
space ($M$ is its point), or, which is equivalent, the function $q(x)$. We
cannot set $q(x)=x$ since $q(x)$ should be periodic. However, we should
define $q(x)$ in such a way that formula 
\[
\psi _{k}(q):=L^{-1/2}\exp [{\rm i}p_{k}x(q)/\hbar ]=L^{-1/2}\exp [{\rm i}%
p_{k}q/\hbar ],
\]%
which is analogous to ({\rm A}.2), be valid. For this to be so $q(x)$ should
only differ from $x$ by periodic jumps of magnitude $\Delta x$ multiple to $L
$ at some points $c_{n}=c_{0}+nL$. If $0<c\leq L/2$, we may set 
$$
q(x)=x-L\eta (x-c)\quad {\rm at}\quad -L/2<x\leq L/2\eqno({\rm A}.3)
$$%
with $\eta (\xi )=(1+{\rm sign}\,\xi )/2$. For function ({\rm A}.3) and $%
\hat{p}=-{\rm i}\hbar \partial /\partial x$ we get 
$$
\lbrack p,q]=-{\rm i}\hbar +{\rm i}\hbar L\delta (x-c)\quad {\rm at}\quad
-L/2<x\leq L/2.\eqno({\rm A}.4)
$$%
Averaging ({\rm A}.4) or, to be exact, the matrix 
$$
\lbrack p,q]_{xx^{\prime }}=-{\rm i}\hbar \lbrack 1-L\delta (x-c)]\delta
(x-x^{\prime })\eqno({\rm A}.5)
$$%
with density matrix $\rho _{x^{\prime }x}$ of the type (2.3) we obtain $%
\langle {\rm i}[p,q]\rangle =0.$ Therefore inequality ({\rm A}.1) for $A=q$, 
$B=p$ gives $\sigma _{q}\sigma _{p}\geq 0.$ So the Heisenberg uncertainty
relations may be violated in our case.

Operator ({\rm A}.5) in the momentum representation is of the form 
$$
\langle p_k|[p,q]|p_l\rangle = -{\rm i}\hbar\delta_{kl} + {\rm i}\hbar
(-1)^{k-l} \eqno ({\rm A}.6) 
$$
in the limit $c\rightarrow L/2$. Therefore $\langle p_k|[p,q]|p_k\rangle =
0. $

Note that the unusual commutativity relation ({\rm A}.5), ({\rm A}.6) leads
to unusual dynamic equations. For example, in the case of an isolated
apparatus with simple Hamiltonian $H_{{\rm A}}=p^{2}/(2m_{0})$ the usual
equation $\dot{q}=p/m_{0}$ is not valid.

\section*{References}

\begin{description}
\item {[1]} {\sc Belavkin, V. P.}, {\it Foundations of Physics}, {\bf 24}%
:685 (1994).

\item {[2]} {\sc von Neumann, J. }, {\it Mathematical Foundations of Quantum
Mechanics}, Princeton University Press, 1955.

\item {[3]} {\sc Stratonovich R. L.}, {\it On the Dynamical Interpretation
for the Collapse of the State during Quantun Measurement}, Proceedings of
the International Conference on Quantum Measurement and Communication,
Nottingham, 1994, Plenum Publisher, 1995.

\item {[4]} {\sc Belavkin, V. P.}, {\it Russian Journal of Mathematical
Physics}, {\bf 3}:3 (1995).

\item {[5]} {\sc Araki, H.}, {\it Progr. Theor. Phys.}, {\bf 64}:719 (1980).

\item {[6]} {\sc Machida, S. and Namiki, H.}, {\it Progr. Theor. Phys.}, 
{\bf 63}:1457 (1980); {\bf 63}:1833 (1980).

\end{description}

\end{document}